\documentstyle[12pt,aaspp4]{article} 
\begin{document}
\title{Fate of the Universe, Age of the Universe, Dark Matter, 
                             and the   Decaying  Vacuum Energy}
\author{Murat \"Ozer}
\affil{ Department of Physics, College of Science, King Saud University
 	         P.O.Box 2455, Riyadh 11451, Saudi Arabia
\footnote{E-mail: mozer@ksu.edu.sa}}
\newcommand{\etal}{{\it et al. }}

\begin{abstract}
It is shown that in cosmological models based on a  vacuum energy  
decaying as $a^{-2}$, where $a$  is the scale factor of the universe,
the fate of 
the universe in regard to whether it will collapse in the  future or 
expand forever is determined not by the curvature constant $k$ but by 
an effective curvature constant $k_{eff}$. It  is argued that a closed 
universe with 
$k=1$ may expand forever, in other words  simulate the expansion dynamics 
of a flat or an open universe because of the possibility that $k_{eff}=0$ 
or -1, respectively. Two such models, in one of which the vacuum does not
interact with matter and in another of which it does, are studied. It is 
shown that the vacuum equation of state $p_{vac}=-\rho_{vac}$ may be 
realized in a decaying vacuum cosmology provided the vacuum interacts 
with matter. The optical depths for gravitational lensing as a function 
of the matter density and other parameters in the models are calculated 
at a 
source redshift of 2. The age of the universe is discussed and shown to 
be compatible with the new {\it Hipparcos} lower limit of $11 Gyr$. The 
possibility that a time-varying vacuum energy may serve as dark matter 
is suggested.
\end{abstract}

(Published in the Astrophysical Journal, 520, 45 (1999))
\section{INTRODUCTION}

Homogeneous and isotropic cosmological models based on a time-varying 
cosmological term $\lambda(t) \propto a(t)^{-2}$, with $a(t)$ being the
scale factor of the universe, do not suffer from the notorious problems
of the standard hot big-bang cosmology such as the initial singularity, 
horizon (causality), entropy, monopole, and cosmological constant 
problem (\cite{oze86}, 1987). However, these are not the only successes 
of such cosmological models. There are others. In the homogeneous and 
isotropic cosmological models based on the Robertson-Walker metric
\begin{equation}%eq.(1)
ds^{2}=-dt^{2}+a(t)^{2}\left[\frac{dr^{2}}{1-kr^{2}}+r^{2}(d\theta^{2}
+sin^{2}\theta d\phi^{2})\right]
\end{equation}
the universe is closed and its expansion will halt and contraction will
 start if the curvature constant $k=1$, the universe is open and will 
expand forever if $k=-1$, and the universe is flat and will expand 
forever if $k=0$. As is well known, these conclusions  concerning the 
expansion of the universe follow from the Friedmann equation
\begin{equation} %eq.(2)
\left(\frac{\dot a}{a}\right)^{2}=\frac{8 \pi G}{3}\rho_{M}(t)+
\frac{\lambda}{3}-\frac{k}{a^{2}},
\label{eqn:friedmann1}
\end{equation}
if the (time-independent) cosmological constant $\lambda$ vanishes. Here 
$\rho_{M}(t)$ is the density of relativistic matter (photons and other 
relativistic particles) in the radiation-dominated era and the density 
of nonrelativistic matter (mainly luminous and nonluminous baryons and 
possibly massive particles such as axions) in the matter-dominated era. 
We have set the speed of light $c$ equal to 1. However, these conclusions
 can drastically change if the cosmological constant $\lambda$ is nonzero.
 Depending on the value of $\lambda$, a closed universe may collapse or 
expand forever for positive $\lambda$, an open or flat universe may 
collapse for negative $\lambda$ (see, for example, \cite{rin77}; 
\cite{lan79}; \cite{fel86}). If, on the other hand, $\lambda$ is allowed 
to vary with time according to $\lambda \propto a^{-2}$ not only some of 
the above features are maintained but the expansion dynamics of a  closed 
universe may be similar to that of a flat or an open universe.

In the standard model (hereafter SM) 
\footnote{Felten \& Isaacman(1986) call the models
 with $\lambda=0$ "standard models". However, we follow the general trend 
in the literature and call the totality of them  the "standard model" and 
refer to each case by its $k$ value (see, for example, \cite{mis70}; 
\cite{wei72}.) Models with zero and nonzero pressure, with or without 
$\lambda$ in both cases, are called the Friedmann models and the 
Lema\^itre models, respectively. The SM with $k=0$ is called the 
Einstein-de Sitter model (\cite{fel86}).}
, the relation between the present value $H_0$  of  Hubble's constant 
$H=\dot a/a$   and the present age $t_0$  of the universe is given by 
(\cite{alr96})
\begin{mathletters} %eq.3a,b,c
\begin{eqnarray}
H_0t_0 &=& 
\frac{1}{(1-\Omega_{M})}\left[1-\frac{\Omega_M}{(1-\Omega_M)^{1/2}}
sinh^{-1}(\Omega_M^{-1}-1)^{1/2}\right]\hspace{2mm},\hspace{2mm}k=-1\\
&=& 2/3 \hspace{2mm}, \hspace{8.5cm} k=0 \\
&=& \frac{1}{(\Omega_M-1)}\left[\frac{\Omega_M}{(\Omega_M-1)^{1/2}}
sin^{-1}(1-\Omega_M^{-1})^{1/2}-1\right] \hspace{2mm},\hspace{2mm}k=1
\end{eqnarray}
\end{mathletters}
where $\Omega_M$ 
\footnote{We will denote the current values of the densities and the
density parameters such as $\rho_{M0}$ and $\Omega_{M0}$ by $\rho_M$ and
$\Omega_M$. The time-dependence of 
variable parameters will be denoted explicitly as in $\rho_M(t)$.} 
is the present value of the matter density parameter    defined as the 
ratio of the present values of the matter energy density to the critical  
energy density 
\begin{equation}%eq.(4)
\Omega_{M}=\frac{\rho_{M}}{\rho_{c}}=\frac{\rho_M}{3H_0^2/8\pi G}
\end{equation}
At this point let us also define $\rho_{\Lambda}=\lambda/8\pi G$, the 
energy density associated with the cosmological constant and 
$\Omega_{\Lambda }=\rho_{\Lambda}/\rho_c$, the fraction of  the present 
critical density contributed today by the  cosmological term. 
Recently, new parallax measurements obtained by the {\it Hipparcos} 
satellite have given the age of the globular clusters, hence the age of 
the universe, as between 11 to 13 Gyr (\cite{rei97}; \cite{fea97}; 
\cite{sch97}). These time scales are considerably lower than the previous 
estimates of $16\pm 2Gyr$ (\cite{pee93}; \cite{cha96}; \cite{sand96}; 
\cite{bol95}) which are in agreement with the SM for $k=-1$  if 
\begin{displaymath}
H_{0}=55\pm 5kms^{-1}Mpc^{-1},
\end{displaymath}
(\cite{san96}; \cite{tam96}) and in conflict  with it if
\begin{displaymath}
H_{0}=73\pm 10kms^{-1}Mpc^{-1},
\end{displaymath}
(\cite{fre97}). Reid (1997) has argued that the {\it Hipparcos} data 
reveal a $7\%$ increase in the distances inferred from the previous 
ground-based data, implying a decrease in $H_0$. He thus concludes that 
the recent Freedman \etal (1997) value of $H_0$ is reduced to 
$H_0=68\pm 9kms^{-1}Mpc^{-1}$. Sandage, however, points out that their 
value of $H_0=55\pm 5kms^{-1}Mpc^{-1}$  is unaltered (Sandage 1997, 
preprint). We should add in passing that ever since its first 
determinations by Hubble, the value of $H_0$ is still one of the most
 controversial parameters of Astronomy. The values found by different 
people using different methods continue to disagree. For example, to 
early March 1996 the lowest value reported is $H_0=
55^{+8}_{-4}kms^{-1}Mpc^{-1}$  (from surface brightness fluctuations) 
whereas the highest one is $H_0=84\pm 4 kms^{-1}Mpc^{-1}$ 
(from luminosity function of planetary nebulae) (\cite{tri96}). To our 
knowledge, the most recent upper bound on $H_0$ has been determined by 
the Supernova Cosmology Project. Using the first seven supernovae at 
$z\geq 0.35$ that have recently been discovered, Kim \etal  (1997) have
 measured the Hubble constant to be $<83kms^{-1}Mpc^{-1}$ and 
$<78kms^{-1}Mpc^{-1}$ in a flat universe with $\Omega_M\geq 0$ and 
$\Omega_M\geq 0.2$, respectively. Writing $H_0=100hkms^{-1}Mpc^{-1}$, 
in our discussions below  we will use $h$ values in the range 0.5 to 0.8. 
Therefore, in the light of the recently determined moderately high values 
of $H_0$ one cannot claim for sure that the age of the universe problem in
 the SM  has disappeared. High values of $H_0$  force the age  to fall 
below the {\it Hipparcos} lower limit of $11 Gyr$. In Figure 1 we present 
the ages in the SM for the three values of $k$. It is seen that for $k=0$
 the ages are between $8$ to $10 Gyr$ if $h>0.6$. 
Now  for low  $\Omega_M$ and $h$  values between 0.5 to 0.6, the predicted 
 ages  are  greater  than $13 Gyr$. As $\Omega_M$  gets larger  the 
favored $h$ values shift towards smaller ones. The problem continues 
to exist if $\Omega_M=0.2$ and $h>0.75$, $\Omega_M=0.3$ to $0.5$ and 
$h\geq 0.75$, $\Omega_M=0.6$ and $h>0.65$, $\Omega_M=0.7$ to $0.8$ and 
$h\geq 0.65$, $\Omega_M=0.9$ to $1.0$ and $h>0.6$.

Long before {\it Hipparcos}, when the estimated ages of the globular 
clusters were in the range $16\pm 2Gyr$ and hence the age problem was 
starker, an immediate solution to this likely problem was provided in the 
1980's by including a (time-independent) cosmological constant $\lambda$   
as in eq.(2) (\cite{pee84}; \cite{blo85}; \cite{kla86}). Another solution 
to this problem was given by Olson \& Jordan (1987) in the framework of a 
time-varying cosmological constant. They showed that in a flat universe 
with $k=0$ ages of the universe old enough to agree with observations 
(in the 1980's) could be obtained with background energy densities of 
the form  $\rho_{b}=\rho_{b0}(a_0/a)^b$, where $b\geq 0$. 

 Still another potential function of a time-varying cosmological constant 
is that  the vacuum energy density   associated with it can be interpreted
 as the density of dark matter. The nature of the dark matter that is 
supposed to exist around galaxies has been another most debated mysteries
 of Astronomy and Cosmology. Many relativistic and nonrelativistic 
particles have been proposed as candidates for dark matter (see the latest 
reviews by Primack (1996) and Srednicki (1996)). 

The purpose of this paper  is to study a phenomenological cosmological
 model based on the vacuum energy density 
$\rho_{\Lambda}(t)=C_1\rho_M+C_2\rho_Ma_0^2/a^2$ 
(see eq.(\ref{eqn:rholambda}) below) in the matter dominated era and show 
that (i) the part of $\rho_{\Lambda}(t)$ decaying as $a^{-2}$ leads to an 
effective curvature constant $k_{eff}$, and it is $k_{eff}$ that governs 
the fate of the universe, (ii) such a $\rho_{\Lambda}(t)$  solves  the age
 problem of the universe, and (iii) $\rho_{\Lambda}(t)$  serves as  dark 
matter. This paper is organized as follows. In section 2 we mention the 
time-varying cosmological constant models briefly, and introduce in 
section 3 the model we study in this paper. We then present our 
calculations under two different assumptions regarding the interaction 
of the vacuum with matter. Section 4 concludes the paper.

\section{THE TIME-VARYING COSMOLOGICAL CONSTANT MODELS}

The time-varying cosmological constant models with 
$\lambda(t)=const.a^{-2}$
were introduced by \"Ozer \& Taha (1986, 1987) in an attempt to solve
the cosmological problems such as the initial singularity, horizon, 
entropy, monopole, and cosmological constant problem (see Weinberg (1989) 
for a review of the cosmological constant problem). The idea of 
\"Ozer \& Taha (1986, 1987)  was then extended  to include a large variety 
of varying cosmological constants  decaying as $a^{-2}$   (\cite{gas87};
 \cite{che90}, 1992; \cite{ber91}; \cite{oze92}; \cite{abd92}; 
\cite{carval92}; \cite{wag93}; \cite{arb94}; \cite{mat95}). In particular,
 it was shown by Chen \& Wu (1990, 1992)  that very general arguments from 
Quantum Cosmology lead to this
form for the effective cosmological constant (see below).

\section{THE MODEL}

Any cosmological model with a cosmological constant is based on the 
observation that  there is an associated energy density $\rho_{\Lambda}=
\lambda/8\pi G$  in terms of which eq.(\ref{eqn:friedmann1}) can be written 
as
\begin{equation} %eq.5
\left(\frac{\dot a}{a}\right)^2=\frac{8\pi G}{3}\left[\rho_M(t)+
\rho_\Lambda(t)\right]-\frac{k}{a^2}.
\label{eqn:friedmann2}
\end{equation}
 The energy density $\rho_{\Lambda}$ is interpreted as the vacuum energy 
density (see e.g. \cite{wei89}). As we have stated above, in this paper we 
shall consider models in which the dependence of the vacuum energy on the
 scale factor $a$ is of the form
\begin{equation} %eq.6
\rho_\Lambda(t)=C_1\rho_M+C_2\rho_M\frac{a_0^2}{a^2},
\label{eqn:rholambda}
\end{equation}				    
where $C_1$ and $C_2$ are constants, $\rho_M$ and $a_0$  are the present 
values of the matter energy density $\rho_M$  and the scale factor $a$. 
The reason for adopting the variable part of $\rho_{\Lambda}\propto a^{-2}$ 
is manyfolds. First, postulating that $\rho_{\Lambda}\propto a^{-n}$ 
$(n\neq 0)$, it is only for $n=2$ that an effective curvature constant 
$k_{eff}$
can be defined. Second, observing that the cosmological constant $\lambda$
has dimension of inverse lenght squared, the simplest scale factor 
dependence of 
$\lambda$ would be $\propto a^{-2}$. This is the case here because 
 the time-varying cosmological constant that corresponds to 
eq.(\ref{eqn:rholambda}) has the form 
\begin{equation} %eq.7
\lambda(t)=\lambda_1+\lambda_2a(t)^{-2},
\label{eqn:lambda}
\end{equation}
where $\lambda_1$ and $\lambda_2$   are constants, and  corresponds to  a 
special case of  the form considered by Matyjasek (1995). Third, on 
dimensional grounds again, $\lambda$
can be written as $\lambda \propto \frac{1}{\ell_{Pl}}
\left(\frac{\ell_{Pl}}{a}\right)^n$, where
$\ell_{Pl}=\left(\frac{\hbar G}{c^3}\right)^{1/2}$ is the Planck lenght.
In a classical theory such as General Relativity an $\hbar$ dependence 
in $\lambda$ is not expected. The correct choice for $n$, therefore, is 2 
(Chen \&Wu 1990; Waga 1993). We have included the constant term in eq.(6)
so as to be able to  compare the variable term with it.

Denoting by $\rho_B$  the sum of luminous and non luminous (dark) baryonic
 energy densities, and by $\rho_{NB}$ the sum of non baryonic energy
 densities the total matter density is $\rho_M=\rho_B+\rho_{NB}$. 
The main argument for baryonic dark matter (see \cite{carr94} for a review)
 is associated with the successful calculations in the standard model of 
the primordial abundances of light elements $\left[X(^4He)\approx 0.24, 
X(^2D)\sim X(^3He)\sim 10^{-5}, X(^7Li)\sim 10^{-10}\right]$. These 
predictions apply only if the baryon density parameter lies in the range 
$0.009h^{-2}\leq \Omega_B\leq 0.02h^{-2}$ (\cite{cop95}; \cite{mal93}, 
\cite{wal91}). On the other hand, the density parameter 
$\Omega_B^{lum}\sim 0.01$   of the luminous baryons is certainly below 
this range. Therefore there must be a significant amount of nonluminous 
baryonic matter in the universe.
 
To proceed further, we need to make an assumption about the interaction of
matter with the variable vacuum energy. There are two possibilities leading
to two distinct models.

\subsection{Model 1}
%sec.3.1 

Matter and the time-dependent vacuum do not interact with each other. 
In this case  the energy conservation equation
\begin{equation} %eq.8
d[\rho_M(t)a^3+\rho_{\Lambda}(t)a^3]+[p_M(t)+p_{\Lambda}(t)]da^3=0
\label{eqn:energy}
\end{equation}
with $p_M(t)=0$   leads to 
\begin{mathletters}
\begin{eqnarray}
d[\rho_M(t)a^3]=0,\\
d[\rho_{\Lambda}(t)a^3]+p_{\Lambda}(t)da^3=0,
\end{eqnarray}
\end{mathletters}
from which  we obtain  $\rho_M(t)=\rho_Ma_0^3/a^3$ and $p_{\Lambda}(t)=
-C_1\rho_{M}-\frac{1}{3}C_2\rho_{M}a_0^2/a^2$ 
\footnote{Stipulating that in a locally inertial frame the vacuum 
energy-momentum tensor be Lorentz invariant requires that it be 
proportional
to the Minkowski metric tensor  $diag(-1, 1, 1, 1, )$ as this is the only 
4x4 matrix that is invariant under Lorentz boosts. On the other hand, the 
energy-momentum tensor of a perfect fluid is of the form 
$diag(\rho, p, p, p)$. Hence vacuum must be a perfect fluid with the 
equation of state $p_{vac}=-\rho_{vac}$. In general, this does not require 
that the vacuum energy and hence the cosmological constant be 
time-independent. $\rho_{vac}$  may be varying with the time. 
To qualify as a vacuum energy it suffices for $\rho_{vac}$    to satisfy 
the above vacuum equation of state. However, in the literature a 
time-varying cosmological term that does not strictly satisfy 
$p_{vac}=-\rho_{vac}$  has also 
been called "time-varying cosmological constant" or "time-varying vacuum 
energy" (\cite{pee88}, \cite{rat88}, \cite{rat92}, 
\cite{pee93}). Our Model 1 does not satisfy the vacuum equation of state 
if $C_2\neq 0$. We shall, however, consider later a model that satisfies 
it as a special case (see Model 2, section 3.2)}. 
The Friedmann equation (\ref{eqn:friedmann2}) then becomes 
\begin{mathletters}%eq.(10a) and (10b)
\begin{eqnarray}
\left(\frac{\dot a}{a}\right)^2&=&\frac{8\pi G}{3}
\rho_{M}\left[\frac{a_0^3}{a^3}+C_1+C_2\frac{a_0^2}{a^2}
\right]-\frac{k}{a^2},\label{eqn:friedmann3a}\\
&=&\frac{8\pi G}{3}\rho_{M}\left[\frac{a_0^3}{a^3}+C_1
\right]-\frac{k_{eff}}{a^2},
\label{eqn:friedmann3b}
\end{eqnarray}
\end{mathletters}
where we have defined
\begin{eqnarray}
k_{eff}&=&k-\frac{8\pi G}{3}C_2\rho_{M}a_0^2 \nonumber \\
&=&k-C_2\Omega_{M}H_0^2a_0^2.
\label{eqn:keff}
\end{eqnarray}
Eq.(\ref{eqn:friedmann3b}) suggests that the fate of the universe in 
regard to whether it will expand forever or collapse  in the future is 
determined  not by $k$, as opposed to the standard model, but by $k_{eff}$.
 This is a rather interesting feature  of the models  with $\lambda(t)$  
varying as $a^{-2}$. Unfortunately, this  feature has not been appreciated 
in the literature well enough. It seems from eq.(\ref{eqn:keff}) that the
 curvature constant  may take on any of the three values 1, 0, and -1. 
If, however, one desires a universe that does not suffer from the initial
 singularity, horizon , and entropy problems, one must then consider the
extension of this model
to the very early universe. There one finds that with a vacuum energy
decaying as $a^{-2}$ the universe does not suffer from these problems only 
if $k=1$ (\cite{oze86}, 1987). Note the intriguing possibility that the 
universe simulates the  expansion dynamics of a flat universe even though 
it is closed ($k=1$). This occurs for $k_{eff}=0$, which is realized if 
\begin{equation}
C_2=\frac{3}{8\pi G\rho_Ma_0^2}=\frac{1}{\Omega_MH_0^2a_0^2}.
\end{equation}
Expressing eq.(\ref{eqn:friedmann3b}) in terms of the present  quantities 
yields  
\begin{equation}
\frac{k_{eff}}{H_0^2a_0^2}=\Omega_M(1+C_1)-1.
\label{eqn:keff-2}
\end{equation}
On the other hand, if $C_2$ is greater than that in eq.(12) $k_{eff}$ will
be negative and the expansion dynamics of the universe will be similar to
that of an open universe.

\subsubsection{ Confrontation of Model 1 with Gravitational Lensing and 
Supernova Studies}

A time-independent cosmological constant has usually been invoked for two 
purposes. First, for large $H_0$, to increase the age of the universe to the
 level of the pre {\it Hipparcos} globular cluster age. Second, to obtain a 
spatially flat universe for low $\Omega_M$, as generally implied by 
inflationary models, so that $\Omega_M+\Omega_{\Lambda}=1$. But does a 
non-vanishing time-independent cosmological constant really exist? Hence it
 is most important to search for ways in which its existence can be tested.
 Fukugita, Futamase \& Kasai (1990) have argued that a statistical  study of
 gravitational lenses could provide for such a test. They have pointed out 
that with a cosmological constant the gravitational lensing optical depth 
(integrated probability) increases very rapidly as the source redshift 
increases. However, they do not make any direct comparison to observational
 data. This has been attempted by Turner (1990). Using the available data on
 the frequency of multiple image lensing of high-redshift quasars by 
galaxies Turner (1990) has shown that spatially flat $k=0$ models with small
 values of $\Omega_M$  and correspondingly large values of 
$\Omega_{\Lambda}$  all predict much larger values of the gravitational 
optical depth. He then concludes that if $k=0$  then $\Omega_M=1$ and 
$\Omega_{\Lambda}=0$ seems to be the favored possibility. Later on, 
Kochanek (1993,1995) and Maoz \& Rix (1993) have managed to put upper bounds
 on using the statistics of lenses. Kochanek (1993, 1995) finds that in a 
flat universe the upper limit on $\Omega_{\Lambda}$ can be as high as 0.8 
 or as low as 0.65. The investigations of Maoz \& Rix (1993) constrain 
$\Omega_{\Lambda}$ to be $\leq 0.7$, also for a spatially flat universe, 
and lead them to conclude robustly that a (time-independent) cosmological 
constant no longer provides an attractive solution for the age problem of 
the universe then. Recently, the supernova magnitude-redshift approach has 
given a value for $\Omega_{\Lambda}$ somewhat smaller than the 
gravitational lens upper limit of Kochanek (1993, 1995). Using the initial 
seven of more than 28 supernovae discovered,  the  Supernova  Cosmology  
Project  has  also  measured $\Omega_M$  and $\Omega_{\Lambda}$ 
(\cite{per97}). They find
$\Omega_M=0.88^{+0.69}_{-0.60}$ for a $\lambda=0$ cosmology, 
 and $\Omega_{M}=0.94^{+0.34}_{-0.28}$ and $\Omega_{\Lambda}=
0.06^{+0.28}_{-0.34}$ 
for a flat universe. They find that $\Omega_{\Lambda}<0.51$  at the 95\% 
confidence level. They, too, conclude that the results for 
$\Omega_{\Lambda}$-versus-$\Omega_M$ are inconsistent with 
$\lambda$-dominated, low density, flat cosmologies that have been proposed 
to reconcile the (pre {\it Hipparcos}) ages of globular cluster stars with 
large Hubble constant values. For the more general case of a 
Friedmann-Lema\^itre cosmology with the sum of $\Omega_M$ and 
$\Omega_{\Lambda}$ unconstrained, they find the lower limit 
$\Omega_{\Lambda }>-2.3$  and the upper limit $\Omega_{\Lambda}<1.1$ for 
$\Omega_M\leq 1$, or $\Omega_{\Lambda}<2.1$ for $\Omega_M\leq 2$. Their 
limits are significantly tighter than the previous limits  of Carroll, 
Press \& Turner (1992).
 
We next examine whether our Model 1 passes the test of  gravitational 
lensing. The integrated 
probability, the so-called optical depth, for lensing by a population of 
singular isothermal spheres of constant comoving density normalized by the 
fiducial case of the SM, the Einstein-de Sitter model, is 
\begin{equation}
P_{lens}=\frac{15}{4}\left[1-\frac{1}{(1+z_s)^{1/2}}\right]^{-3}
\int_0^{z_s}\frac{(1+z)^2}{E(z)}\left[\frac{d(0, z)d(z, z_s)}{d(0, z_s)}
\right]^2dz
\label{eqn:plens}
\end{equation}
(\cite{carrol92})
\footnote{It is assumed in eq.(\ref{eqn:plens}) that the present 
matter pressure is negligible; hence $\rho_M(t)=\rho_{M}a_0^3/a^3$. It does
not hold in models with nonzero matter pressure at present.}, where 
\begin{equation}
E(z)^2=(1+z)^2(1+z\Omega_{M})-z(z+2)\Omega_{\Lambda}
\label{eqn:e(z)}
\end{equation}
and is defined by
\begin{equation}
\left(\frac{\dot a}{a}\right)^2=H_0^2E(z)^2
\end{equation}
(\cite{pee93}). Note that $P_{lens}=1$ for the Einstein-de-Sitter model 
(in which $\Omega_{M}=1$, $\Omega_{\Lambda}=0$)
. $z=(a_0/a)-1$ is the redshift and $z_s$ is the redshift of the source 
(quasar). The angular diameter distance from redshift $z_1$ to redshift 
$z_2$ is given by 
\begin{equation}
d(z_1,z_2)=\frac{1}{(1+z_2)\mid\Omega_{k}\mid^{1/2}}sinn\left[\mid
\Omega_{k}\mid^{1/2}\int_{z_1}^{z_2}\frac{dz}{E(z)}\right]
\label{eqn:d}
\end{equation}
where $\Omega_{k}=-k/(H_0^2a_0^2)$ and "$sinn$" is defined as $sinh$ if
 $\Omega_{k}>0$, as $sin$ if $\Omega_{k}<0$ and as unity if $\Omega_{k}=0$ 
in which case the $\mid\Omega_k\mid^{1/2}$'s disappear from 
eq.(\ref{eqn:d}).  Equation (\ref{eqn:friedmann3a}) or 
(\ref{eqn:friedmann3b}) gives
\begin{equation}
E(z)^2=(1+z)^2(1+z\Omega_{M})-z(z+2)C_1\Omega_{M},
\label{eqn:e(z)-1}
\end{equation}
where we have used the constraint
\begin{eqnarray}
\Omega_{M}+C_1\Omega_{M}+C_2\Omega_{M}+\Omega_{k}&=&1 \nonumber\\
\Omega_{M}+C_1\Omega_{M}+\Omega_{k_{eff}}&=&1
\label{eqn:sumofomegas} 
\end{eqnarray}
 to eliminate $\Omega_{k}$. It is thus seen upon comparing 
eq.(\ref{eqn:e(z)-1}) with eq.(\ref{eqn:e(z)}) that we only need to 
replace $\Omega_{\Lambda}$ with $\Omega_{C_1}=C_1\Omega_{M}$ to convert 
the expressions (\ref{eqn:plens}) and (\ref{eqn:d}) to this model.We 
present in Table 1 the normalized optical depths in the Friedmann model 
with $k=0$ for 
a typical source redshift of $z_{s}=2$.

Taking the upper bound on $\Omega_{\Lambda}$ as 0.5 (\cite{per97}) we see 
that the corresponding lensing prediction is $P_{lens}=1.92$. Hence, we 
shall assume in the following that the maximum tolerable  value is about 2. 
The predictions increase very rapidly for larger source redshifts.  
We present a sample from our predictions
 for $k=1$ and $k_{eff}=0$ in Figure 2 subject to the constraint equation 
(\ref{eqn:sumofomegas}) from which it follows in this case that 
$C_{1}=1/\Omega_{M}-1$ and $C_{2}=-\Omega_{k}/\Omega_{M}$. It is seen that
 plausible $P_{lens}$ values 
are obtained only for $\Omega_{M} \geq 0.5$, or equivalently for 
$\Omega_{C_{1}}\leq 0.5$ with the corresponding $C_{1}$ values being 
$\leq 1$ only.
				  			     
\subsubsection{ Age of the Universe in Model 1}
%sec.3.1.2
Inserting eq.(\ref{eqn:keff-2}) in eq.(\ref{eqn:friedmann3b})  we obtain 
the 
relation between the present age $t_0$ of the  universe    and the present
 value of the Hubble's constant $H_0$:
\begin{equation}
H_0t_0=\Omega_{M}^{-1/2}\int_0^{1}y^{1/2}\{1+C_1y^3+
[\Omega_{M}^{-1}-(1+C_1)]y\}^{-1/2}dy.
\label{eqn:H0t0-1}
\end{equation}
It is worth noting that this expression is independent of $C_2$, namely 
the age of the universe in  Model 1 is independent of the part of 
$\rho_{\Lambda}$ varying as $a^{-2}$  (However, see Model 2 below). 
For $k_{eff}=0$ ($\Omega_{M}+C_1\Omega_{M}=1$) eq.(\ref{eqn:H0t0-1}) 
reduces to
\begin{equation}
H_0t_0(k_{eff}=0)=\frac{2}{3(1-\Omega_{M})^{1/2}}\sinh^{-1}
(\Omega_{M}^{-1}-1)^{1/2},
\label{eqn:H0t0-2}
\end{equation}
which is identical to that in a universe with $k=0$. With  
$H_0=100hkms^{-1}Mpc^{-1}$ the age is  $t_0=(9.78/h)(H_0t_0)Gyr$, 
where $H_0t_0$  is calculated from equations (\ref{eqn:H0t0-1}) or 
(\ref{eqn:H0t0-2}), depending on the value of $k_{eff}$ . 

Next we address ourselves the question of whether  the ages in the SM that
are below the {\it Hipparcos} lower limit of $11 Gyr$ in Figure 1 could be
raised to $11 Gyr$ with the help of a decaying cosmological term as 
considered in this section. Starting with $h=0.60$, we have determined the 
value of $C_1$, by trial and error from eq.(\ref{eqn:H0t0-1}), that gives 
the age as $11 Gyr$ for a certain $\Omega_{M}$. Then we have calculated the
corresponding lensing prediction as a function of $C_2$. The values thus 
obtained are presented in Figure 3 for $h=0.60$ and $0.65$ and in Figure 4
for $h=0.70$ and $0.80$. Disqualifying $P_{lens}$ values that are 
significantly over 2, as we have 
decided before, it is seen from Figure 4 that $h=0.70$ and 
$\Omega_{M}\geq 0.9$, and $h=0.80$ and $\Omega_{M}>0.2$  are unsuccessful. 
The time-independent component of $\rho_{\Lambda}$  cannot help increase 
the age for the troublesome values of $h$ and $\Omega_{M}$.

Even though the range of $\Omega_{M}$  was extended  up to  2 in Figure 1, 
we have not extended its range beyond 1  in our other calculations  because 
the current estimates usually give a value between 0.1 to 1. For example,
 early dynamical estimates of the clustered mass density suggested 
$\Omega_{M}=0.2\pm 0.1$ (\cite{pee86}; \cite{bro87}). Recent studies of 
 galaxy clusters give $\Omega_{M}=0.19\pm 0.06$(\cite{carl97}), which is 
comparable to the least action principle result of 
$\Omega_{M}=0.17\pm 0.10$ (\cite{sha95}). On the other hand, the methods 
that sample the largest scales via peculiar velocities of galaxies and 
their production through potential fluctuations yield values of 
$\Omega_{M}$ close to unity (\cite{dek96}).
	
\subsubsection{ Dark Matter from the Decaying Vacuum Energy} 

Before we consider the implications of a different assumption for the 
interaction of  matter and the vacuum next, we  note that equation 
(\ref{eqn:sumofomegas}) suggests that the decaying part of 
$\rho_{\Lambda}$
can be interpreted as the energy density of dark matter (nonluminous matter) 
with
\begin{equation}
\Omega_{DM}=C_2\Omega_{M}.
\label{eqn:dm}
\end{equation}
This is similar to the dark matter from a homogeneous scalar field 
(\cite{rat88}; \cite{pee93}). Note also that part of $\Omega_{M}$ may 
actually be due to the  decaying part of $\rho_{\Lambda}$. Since there is 
no way of knowing this we have used the observational limits for 
$\Omega_{M}$ in our calculations. This will not change our conclusions in 
any way. If the ideas we propose here are correct there must be dark matter
 due to $\rho_{\Lambda}$ not only around the galaxies but also in between 
the galaxies. Therefore we predict much more dark matter especially in 
between the galaxies than  there is around the galaxies. Thus, despite the 
fact that the $C_1$ part of $\rho_{\Lambda}$ cannot offer a satisfactory 
solution to the new age problem, it can help close the universe, even though
 by a small amount allowed by gravitational lensing, together with the 
$C_2$ part so that $k=1$ provided $\Omega_M (1+C_1+C_2)>1$, or produce a 
$k_{eff}=0$ or -1 universe. We have also checked that a vanishing or a 
negative $C_1$  gives better lensing predictions. But negative values of 
$C_1$ not only  necessitate higher values of $C_2$, hence more dark matter 
would be required to close the universe, but also aggravate the age problem.
A purely constant term due to a relic time-independent cosmological 
constant, like the $C_1$  component of $\rho_{\Lambda}$, was considered 
previously by Peebles (1984) and Turner, Steigman \& Krauss (1984) to obtain
a flat $k=0$ universe. Such a term is spatially constant and cannot be 
considered dark matter, even though it modifies the total energy density 
parameter $\Omega_T=\Omega_M+\Omega_{\Lambda}$. A spatially nonuniform 
energy density, however, can serve as dark matter. 
		
\subsection{ Model 2} 

Matter and the time-dependent vacuum interact with each other as governed 
by the equation 
\begin{equation}
d[\rho_M(t)a^3]+d[\rho_{\Lambda}(t)a^3]+w\rho_{\Lambda}(t)da^3=0
\label{eqn:energy-2}
\end{equation}
where $p_M(t)=0$ and $p_{\Lambda}(t)=w\rho_{\Lambda}(t)$. Here $w$ is a 
negative parameter that is greater than or equal to - 1. The case with 
$w=-1$ is of great interest in that it shows that the vacuum equation of 
state may be realized with a decaying cosmological term. This is rendered 
possible due to the interaction of the vacuum with matter and is not 
allowed, for instance, in our Model 1 in which the vacuum and matter do 
not interact. Substituting eq.(\ref{eqn:rholambda}) into 
eq.(\ref{eqn:energy-2})  yields
\begin{equation} %eq.(24)
\rho_M(t)=[1+(w+1)C_1+(3w+1)C_2]\rho_M\frac{a_0^3}{a^3}-(3w+1)C_2
\rho_M\frac{a_0^2}{a^2}-(w+1)C_1\rho_M,
\end{equation}
where  the $a^{-2}$  term is due to the decay of the vacuum into matter. 
At this point we must ask 'What particles does the vacuum decay into?'. 
One can only speculate. The vacuum may be decaying into various forms of 
matter such as baryons and axions. The Friedmann equation 
(\ref{eqn:friedmann2}) now becomes
\begin{equation}
\left(\frac{\dot a}{a}\right)^2=\frac{8\pi G}{3}\rho_{M}\left\{[1+(w+1)C_1+
(3w+1)C_2]\frac{a_0^3}{a^3}-wC_1\right\},
\label{eqn:friedmann4}
\end{equation}
where now 
\begin{eqnarray}
k_{eff}&=&k-8\pi G\mid w\mid C_2\rho_{M}a_0^2 \nonumber\\
&=&k-3\mid w\mid C_2\Omega_{M}H_0^2a_0^2,
\end{eqnarray}
and	
\begin{equation}
\frac{k_{eff}}{H_0^2a_0^2}=\Omega_{M}[1+C_1+(3w+1)C_2]-1.
\end{equation}
Having presented the salient features of this model we next test it against 
gravitational lensing.
	
\subsubsection{ Gravitational Lensing and Age of the Universe in Model 2}

It follows from eq.(\ref{eqn:friedmann4}) that $E(z)^2$ is now given by 
\begin{equation}
E(z)^2=(1+z)^2(1+z\Omega_{{M}_{eff}})-z(z+2)\Omega_{C_1},
\label{eqn:e(z)-2}
\end{equation}				
where
\begin{eqnarray}	
\Omega_{{M}_{eff}}&=&[1+(w+1)C_1+(3w+1)C_2]\Omega_{M},\nonumber \\
\Omega_{C_1}&=&-wC_1\Omega_{M}.
\label{eqn:omegaeff}
\end{eqnarray}	    
This time, equations (\ref{eqn:plens}) and (\ref{eqn:d}) are converted to 
the present model under the replacement $(\Omega_{M},\Omega_{\Lambda},
\Omega_{k})\rightarrow (\Omega_{{M}_{eff}},\Omega_{C_1},\Omega_{k})$. 
Hence,  the two set of parameters having the same values will yield the 
same lensing predictions. In this case, a closed universe will have the 
same expansion dynamics of a flat one if $\Omega_{{k}_{eff}}=
1-[1+C_1+(3w+1)C_2]\Omega_{M}=0$, which is realized if 
\begin{equation}
C_2=1/(\mid w\mid 8\pi G\rho_{M}a_0^2)=1/(3\mid w\mid \Omega_{M}H_0^2a_0^2).
\label{eqn:C2}
\end{equation}
The relation  between  $t_0$ and  $H_0$  now is 
\begin{eqnarray}
H_0t_0=\Omega_{M}^{-1/2}\int_0^1 y^{1/2} \{1+(w+1)C_1+
(3w+1)C_2-wC_1y^3\nonumber \\	         
\hspace{4cm}+[\Omega_{M}^{-1}-1-C_1-(3w+1)C_2]y \}^{-1/2}.
\label{eqn:H0t0-3}
\end{eqnarray}								
An investigation of the integrand  in eq.(\ref{eqn:H0t0-3}) shows that for 
the integral to be real valued  $(w+1)C_1+(3w+1)C_2>-1$.

This model has an interesting property not shared by the previous one. For 
$k_{eff}=0$ and $C_1=0$ it has $\Omega_{M_{eff}}=1$, and hence the same 
prediction for the age of the universe as the fiducial case  of the SM, 
the Einstein-de Sitter model. This is true  for all values of $\Omega_{M}$
 satisfying $\Omega_{M}+(1/3\mid w\mid -1)/(H_0^2a_0^2)=1$, which follows 
from equations (\ref{eqn:omegaeff}) and (\ref{eqn:C2}). It is of interest 
to note that if $\mid w\mid <1/3$ a universe with $\Omega_{M}<1$ and 
$k_{eff}=0$ remains a possibility in this model with $C_1=0$.
(If $\mid w\mid =1/3$ it is necessary that $\Omega_{M}=1$ so that 
$k_{eff}=0$.) The age of the universe in this case is then $t_0=
(6.52/h)Gyr$ (see eq.(3b)). Thus this case can survive only if  the 
parameter $h$ is
$\leq 0.60$  (see Figure 1). For $k_{eff}=0$  eq.(\ref{eqn:H0t0-3}) 
reduces to 
\begin{eqnarray}
H_0t_0(k_{eff}=0)&=&\frac{2}{3(\mid w\mid C_1\Omega_M)^{1/2}}\sinh^{-1}
\left(\frac{\mid w\mid C_1\Omega_{M}}{1-\mid w\mid C_1\Omega_{M}}
\right)^{1/2} \nonumber \\	
&=&\frac{2}{3(1-\Omega_{{M}_{eff}})^{1/2}}\sinh^{-1}
(\Omega_{{M}_{eff}}^{-1}-1)^{1/2}
\end{eqnarray}	    		     
for positive $C_1$, and to 
\begin{eqnarray}
H_0t_0(k_{eff}=0)&=&\frac{2}{3(\mid w\mid C_1\Omega_{M})^{1/2}}\sin^{-1}
\left(\frac{\mid w\mid C_1\Omega_{M}}{1+\mid w\mid C_1\Omega_{M}}
\right)^{1/2} \nonumber \\	
&=&\frac{2}{3(1-\Omega_{{M}_{eff}})^{1/2}}\sin^{-1}
(\Omega_{{M}_{eff}}^{-1}-1)^{1/2}
\label{eqn:negC1}
\end{eqnarray}	
for negative $C_1$. We see upon examining eq.(\ref{eqn:negC1}) that the 
possibility of negative $C_1$ aggravates the age problem very severely.
The special case with $C_1=0$ offers further possibilities. For 
$\Omega_{{M}_{eff}}<1$ eq.(\ref{eqn:H0t0-3}) reduces to
\begin{eqnarray}
\label{eqn:H0t0-4}
H_0t_0&=&\frac{1}{\{1-[1+(3w+1)C_2]\Omega_M\}} \nonumber \\
       &\times &\left[1-\frac{[1+(3w+1)C_2]\Omega_{M}}{\{1-[1+(3w+1)C_2]
\Omega_{M}\}^{1/2}}\sinh^{-1}\{[1+(3w+1)C_2]
\Omega_{M}]^{-1}-1\}^{1/2}\right],\nonumber \\
&=&\frac{1}{(1-\Omega_{{M}_{eff}})^{1/2}}\left[1-\frac{\Omega_{{M}_{eff}}}
{(1-\Omega_{{M}_{eff}})^{1/2}}\sinh^{-1}(\Omega_{{M}_{eff}}-1)^{1/2}\right].
\end{eqnarray}
As expected, this is similar to the SM expression for $k=-1$, eq.($3a$), 
and $k_{eff}$ is negative as long as $C_1=0$ and $\Omega_{{M}_{eff}}<1$. 
In particular, for $w=-1/3$ and $\Omega_M<1$ equations (\ref{eqn:H0t0-3})
 and (\ref{eqn:H0t0-4}) reduce to eq.($3a$). For $w=-1/3$ and 
$\Omega_{{M}_{eff}}>1$ $k_{eff}$ is positive and eq.(\ref{eqn:H0t0-3}) 
reduces to eq.($3c$), which is the $k=1$ case of the SM. Once again we 
emphasize that it is $k_{eff}$ but not $k$ that determines the expansion 
fate of the universe. In Table 2 we show the values of $C_2$ that are 
required to raise the age to $11 Gyr$ against $\Omega_M$ and $h$  
for $C_1=0$ and $w=-1$. The lensing predictions now are very promising and 
it seems that values of $h$  greater than 0.8 may be allowed. As can be seen
 from eq.(\ref{eqn:H0t0-4}) that the maximum value of $H_0t_0$ when $C_1=0$ 
 for negative $k_{eff}$ is unity, in which case the age of the universe is 
 $9.78/h Gyr$. The noteworthy point here is that the $H_0t_0$ values near 
one are obtained for all values of $\Omega_M$(see Figure 5), whereas in 
the SM a value for $H_0t_0$  as large as unity can only be obtained as
 $\Omega_M$ tends to zero for $k=-1$ (see eq.($3a$). 

In Tables 3, 4 and 5 we display the $H_0t_0$ and $P_{lens}$ predictions for 
$w=-1$, $-2/3$ and $-1/3$, respectively. A value of $C_2$ near $1$ increases
 the age by $10$ to $25-30$ percent over the SM as $\Omega_M$ increases 
from $0.1$ to $1.0$, while $P_{lens}$ remains within the acceptable range 
for $w=-1$ and $-1/3$ but increases slightly over 2 for $w=-2/3$. The cases
 with $w=-1$ in Tables 2 and 3 are particularly interesting because they 
have a strictly vacuum equation of state, i.e. $p_{\Lambda}=-\rho_{\Lambda}$. 

Finally we note that an independent estimate on $C_1+C_2$ can be obtained 
as follows: Comparing eq.(\ref{eqn:rholambda}) with eq.(\ref{eqn:lambda}) 
yields
\begin{equation}
\lambda =(C_1+C_2)\frac{8\pi G}{c^2}\rho_M,
\end{equation}
where we have retained the speed of light $c$. Dividing through by 
$3H_0^2$ we 
get
\begin{equation}
\Omega_{\Lambda}=(C_1+C_2)\Omega_M,
\end{equation}
which is consistent with  $C_1+C_2\leq 1$.

\section{ DISCUSSION AND  CONCLUSIONS}

The ongoing recent research in cosmology and related fields reveals that it
 is still taken for granted  by most people that the cosmological constant 
is a true constant which does not change with time. We have shown in our 
phenomenological model here that the vacuum equation of state 
$p_{vac}=-\rho_{vac}$  may be 
realized with a time-dependent cosmological term (see also \cite{carval92}).
 A truly time-independent cosmological constant itself introduces a very 
serious problem, the so called cosmological constant problem (\cite{wei89}),
 to which no satisfactory solution has been found so far. In view of the 
cosmological constant problem and the suggested solution to it 
(\cite{oze86}, 1987) we cannot think of a mechanism to reduce the constant 
term in eq.(\ref{eqn:rholambda}) from its very large value in the early 
universe to its 
observational limits today. Gravitational lensing statistics limit the value
 of such a constant term to a level that it cannot help increase the age of 
the universe to the {\it Hipparcos} lower limit of $11 Gyr$ for all $h$ up 
to 0.8. If 
the present value of the matter density parameter is low enough neither can 
it help to have a $k=0$ universe in the Friedmann models. Thus it just seems 
redundant
. We, therefore, believe very strongly that there should be no such constant
 component  in $\rho_{\Lambda}$.

It is only natural to associate a temperature with the vacuum energy. 
Assuming that the vacuum, which had enormous energy in the early universe, 
was in thermal equilibrium with the matter content of the universe at least 
during some period in the past, it is necessary that the temperature of the
 vacuum decreases with the expansion of the universe (\cite{gas87}, 1988). 
Thus a decaying vacuum energy or equivalently a decaying cosmological 
constant  seems to be more plausible than a time-independent one. We have 
shown in this work that, a decaying cosmological constant can increase  the 
 age to 11 Gyr without running into conflicts with the gravitational lensing
 predictions for all $h\leq 0.80$, and even for larger values provided the 
vacuum and 
matter interact with each other (as in our Model 2). We have also argued 
here that a decaying cosmological constant can serve as dark matter. A very 
intriguing possibility is that a cosmological constant decaying as  
$a^{-2}$
necessitates the definition of an effective curvature constant $k_{eff}$, 
which in 
turn may cause a closed universe (with $k=1$) to have similar expansion 
dynamics as a  flat universe if $k_{eff}=0$ or as an open universe 
if $k_{eff}=-1$. 
In our opinion, this is an intriguing possibility of such cosmologies
 that deserve further study. However, these are not the only successes of  
such cosmologies. We   should like to mention that these models do not 
suffer from cosmological problems such as the initial singularity, horizon,
 and  entropy  problems of the standard model and thus yield a problem free 
universe (\cite{oze86}, 1987).

The interesting and ultimate question is to explain how a   time-dependent 
 term, and in particular one decaying as $a^{-2}$ can arise in the first 
place. 
A first attempt in this direction has been undertaken by 
Peebles \& Ratra (1988) and Ratra \& Peebles (1988). They have considered a 
model in which the vacuum energy  $\rho_{\Lambda}$ depends on a scalar field
 that changes as
 the universe expands (see also \cite{rat92}). In the two string-motivated 
scalar-field  cosmological models of  \"Ozer \& Taha (1992) the universe is 
closed and non-singular. The scalar fields have a negative pressure  of 
$-\frac{1}{3}\rho_{\phi}$ and $-\frac{2}{3}\rho_{\phi}$   
in these models. Furthermore, the energy density $\rho_{\phi}$  of the 
scalar field 
decays like $a^{-2}$. Hence these scalar field models mimic the decaying 
cosmological  constant models considered here. This endeavor is currently 
under further investigation. Other objects whose energy densities decrease
as $a^{-2}$ are the textures, which are topological defects. The most studied
such objects are the non-Abelian cosmic strings (\cite{vil93}). If strings
do not intercommute nor pass through each other, then their energy density 
scale as $a^{-2}$ (\cite{vil84}). Recently, a closed universe with 
$\Omega_M<1$ and some form of matter (a scalar field, a cosmic string or 
some other stable texture) with an equation of state $p=-\rho/3$ and energy 
density $\rho$ scaling as $a^{-2}$ was considered by Kamionkowski \& Toumbas
 (1996). Another work which is also similar to the present one is the work of
Spergel \& Pen (1996) who considered a flat universe dominated at present 
by cosmic strings. The universe in these two models locally resemble an open
universe, namely $k_{eff}=-1$ even though $k=1$ in the first and 0 in the 
second one. These authors argue that such cosmological models are currently 
viable and  thus represent alternatives to the SM.

In conclusion, phenomenological cosmological models or their field theoretic
 partners based on a vacuum energy of the form 
$\rho_{\Lambda}(t)=C_1\rho_M+C_2\rho_Ma_0^2/a^2$, preferably 
with $C_1=0$, can raise
 the age of the universe up to the {\it Hipparcos} lower limit of $11 Gyr$
 and serve as dark matter. When $C_1=0$ the maximum value of $H_0t_0$
  is equal to one 
corresponding to $t_0=(9.78/h)Gyr$, and there is no problem with the 
age even if  $h$ is as large as 0.85.

We have not attempted in this paper to study the growth of 
inhomogeneities in cosmological models of the type considered here. We hope 
to undertake this endeavor in future work.

ACKNOWLEDGEMENTS

We acknowledge invaluable discussions with Prof. Mahjoob O. Taha. We wish 
to thank 
Prof. Edward L. Wright for informing us of {\it Hipparcos}, the work of 
Perlmutter et al. (1997), suggestions, and pointing out  an error in the 
manuscript.

\clearpage
\begin{deluxetable}{rrr}
\tablecaption{Normalized optical depths in a $k=0$ universe.}
\label{table:1}
\tablewidth{0pt}
\tablehead{
\colhead{$\Omega_{M}$} &\colhead{$\Omega_{\Lambda}$} &\colhead{$P_{lens}$}}
\startdata
  0 &1.0 &13.25\nl
0.1 &0.9 &5.98\nl
0.2 &0.8 &3.94\nl
0.3 &0.7 &2.93\nl
0.4 &0.6 &2.33\nl
0.5 &0.5 &1.92\nl
0.6 &0.4 &1.63\nl
0.7 &0.3 &1.42\nl
0.8 &0.2 &1.25\nl
0.9 &0.1 &1.11\nl
1.0 &0   &1.00\nl
\enddata
\end{deluxetable}
%\clearpage
%
\begin{deluxetable}{cccccc}
\tablecaption{Values of $C_2$ to raise the age to $11 Gyr$ in Model 2
\tablenotemark{a,b}}
\label{table:2}
\tablewidth{0pt}
\tablehead{
\colhead{} &\colhead{} &\colhead{} &\colhead{$h$} &\colhead{} &\colhead{} 
} 
\startdata
 $\Omega_{M}$ &0.8 &0.75 &0.70 &0.65 &0.60\nl
\tableline
0.1 &0.01(1.71) & & & &\nl
0.2 &0.26(1.75) & & & &\nl
0.3 &0.15(1.59) &0.15(1.59) & & &\nl
0.4 &0.38(1.83) &0.24(1.63) & & &\nl
0.5 &0.40(1.87) &0.30(1.68) &0.14(1.45) & &\nl
0.6 &0.42(1.92) &0.33(1.71) &0.19(1.46) & &\nl
0.7 &0.43(1.96) &0.35(1.74) &0.24(1.50) &0.07(1.24) &\nl
0.8 &0.44(2.01) &0.37(1.78) &0.27(1.52) &0.13(1.27) &\nl
0.9 &0.44(2.04) &0.39(1.84) &0.30(1.57) &0.17(1.29) &\nl
1.0 &0.45(2.11) &0.40(1.88) &0.32(1.60) &0.20(1.31) &0.03(1.04)\nl
\enddata
\tablenotetext{a}{$C_1=0$, $w=-1$}  
\tablenotetext{b}{The numbers in parentheses are the corresponding 
$P_{lens}$ values}
\end{deluxetable}
\begin{deluxetable}{cccc}
\tablecaption{Ages and normalized optical depths for $w=-1$ in Model 2
\tablenotemark{a} }
\label{table:3}
\tablewidth{0pt}
\tablehead{
\colhead{$\Omega_{M}$}  &\colhead{$\Omega_{k}$} 
&\colhead{$H_0t_0$}  &\colhead{$P_{lens}$} 
} 
\startdata

0.1 &0.855  &0.98 &1.89\nl
0.2 &0.71   &0.97 &1.91\nl
0.3 &0.565  &0.95 &1.93\nl
0.4 &0.42   &0.94 &1.95\nl
0.5 &0.275  &0.94 &1.98\nl
0.6 &0.13   &0.93 &2.00\nl
0.7 &-0.015 &0.92 &2.03\nl
0.8 &-0.16  &0.91 &2.05\nl
0.9 &-0.305 &0.90 &2.08\nl
1.0 &-0.45  &0.90 &2.11\nl
\enddata
\tablenotetext{a}{$C_1=0$, $C_2=0.45$}  
\end{deluxetable}
%
%\clearpage
\begin{deluxetable}{cccc}
\tablecaption{Ages and normalized optical depths for $w=-2/3$ in Model 2
\tablenotemark{a} }
\label{table:4}
\tablewidth{0pt}
\tablehead{
\colhead{$\Omega_{M}$}  
&\colhead{$\Omega_{k}$\tablenotemark{b}}  
&\colhead{$H_0t_0$\tablenotemark{b}}  &\colhead{$P_{lens}$\tablenotemark{b}} 
} 
\startdata

0.1 &0.855,  0.805   &0.93, 0.99   &1.80, 1.91\nl
0.2 &0.710,  0.610   &0.89, 0.98   &1.74, 1.96\nl
0.3 &0.565,  0.415   &0.86, 0.97   &1.68, 2.01\nl
0.4 &0.420,  0.220   &0.84, 0.97   &1.63, 2.07\nl
0.5 &0.275,  0.025   &0.82, 0.96   &1.59, 2.12\nl
0.6 &0.130,  -0.170  &0.80, 0.95   &1.54, 2.18\nl
0.7 &-0.015, -0.365  &0.78, 0.95   &1.50, 2.24\nl
0.8 &-0.160, -0.560  &0.77, 0.94   &1.47, 2.31\nl
0.9 &-0.305, -0.755  &0.75, 0.94   &1.43, 2.38\nl
1.0 &-0.450, -0.950  &0.74, 0.94   &1.40, 2.45\nl
\enddata
\tablenotetext{a}{$C_1=0$}  
\tablenotetext{b}{The first and second numbers are for $C_2=0.45$ and 
$0.95$, respectively.}
\end{deluxetable}
%\clearpage
\begin{deluxetable}{cccc}
\tablecaption{Ages and normalized optical depths for $w=-1/3$ in Model 2
\tablenotemark{a} }
\label{table:5}
\tablewidth{0pt}
\tablehead{
\colhead{$\Omega_{M}$} 
&\colhead{$\Omega_{k}$\tablenotemark{b}} &\colhead{$H_0t_0$}  
&\colhead{$P_{lens}$\tablenotemark{b}} 
} 
\startdata

0.1 &0.855,  0.805   &0.90   &1.72, 1.73\nl
0.2 &0.710,  0.610   &0.85   &1.60, 1.62\nl
0.3 &0.565,  0.415   &0.81   &1.49, 1.52\nl
0.4 &0.420,  0.220   &0.78   &1.40, 1.44\nl
0.5 &0.275,  0.025   &0.75   &1.33, 1.37\nl
0.6 &0.130, -0.170   &0.73   &1.26, 1.30\nl
0.7 &-0.015, -0.365  &0.71   &1.20, 1.25\nl
0.8 &-0.160, -0.560  &0.70   &1.14, 1.20\nl
0.9 &-0.305, -0.755  &0.68   &1.09, 1.15\nl
1.0 &-0.450, -0.950  &0.67   &1.05, 1.11\nl
\enddata
\tablenotetext{a}{$C_1=0$}  
\tablenotetext{b}{The first and second numbers are for $C_2=0.45$ and 
$0.95$, respectively.}
\end{deluxetable}
\clearpage
%\figcaption[fig1.eps]
%\figcaption[fig2.eps]
%\figcaption[fig3.eps]
%\figcaption[fig4.eps]
%\figcaption[fig5.eps]
%
\begin{figure}
\epsscale{.8}
\plotone{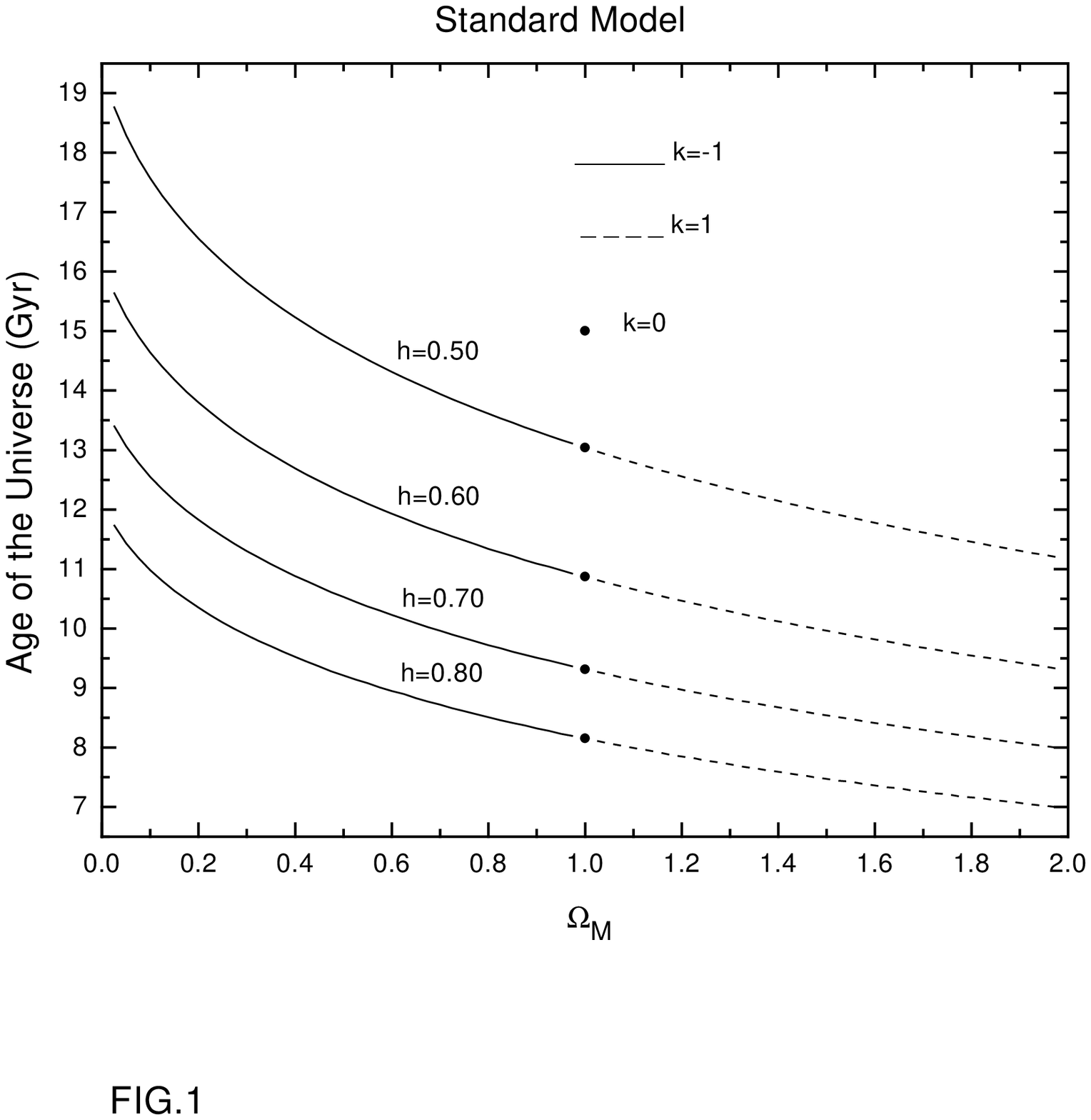}
\caption{The age of the universe in the SM for $k=-1$ 
(solid lines), $k=0$ (dots) and $k=1$ (dashed lines) versus the present 
value of the matter density parameter $\Omega_M$.}
\end{figure}
\begin{figure}
\epsscale{.8}
\plotone{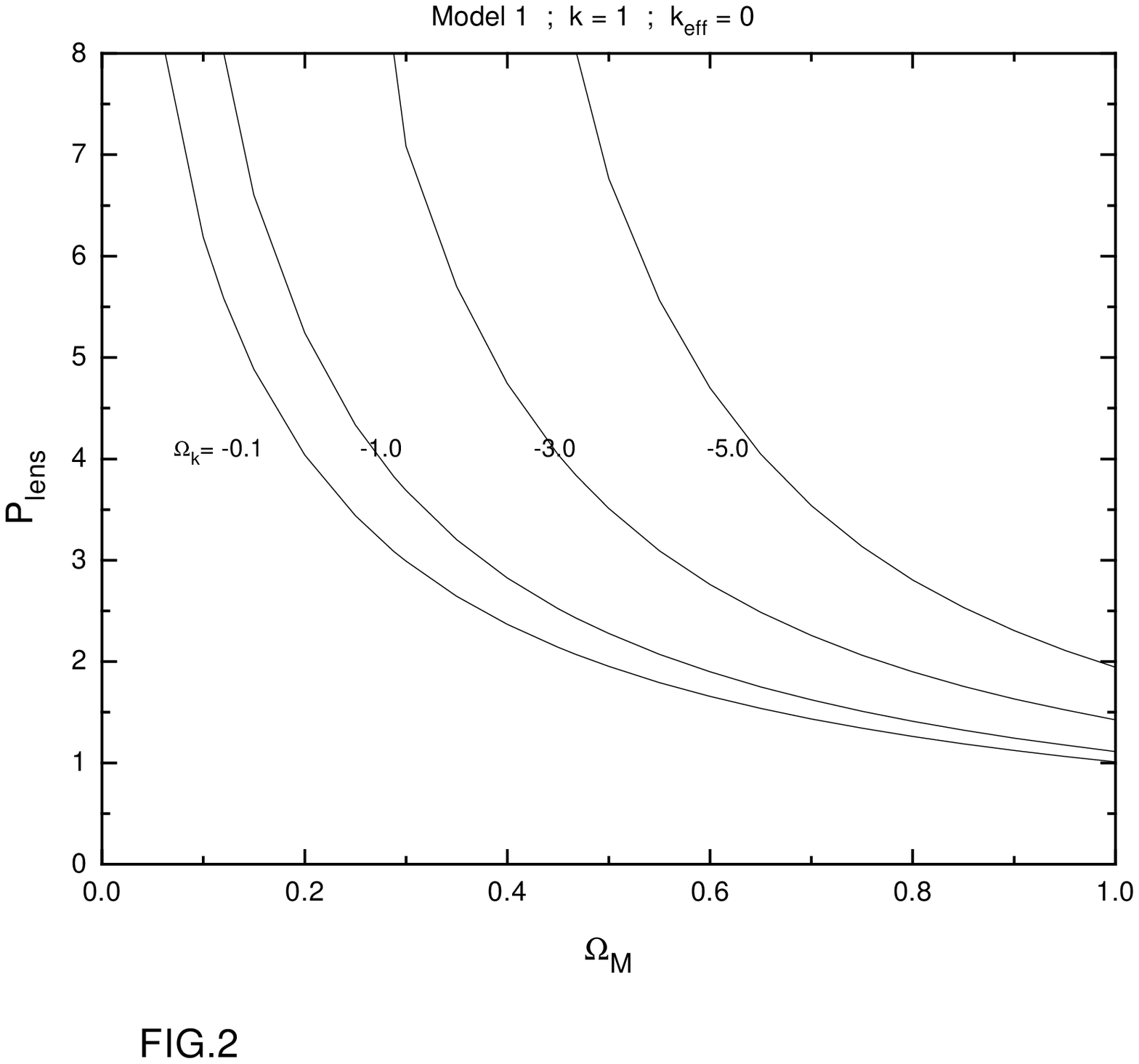}
\caption{Plot of $P_{lens}$ versus $\Omega_M$ in Model 1
for various values of $\Omega_k$.}
\end{figure}
\begin{figure}
\epsscale{.8}
\plotone{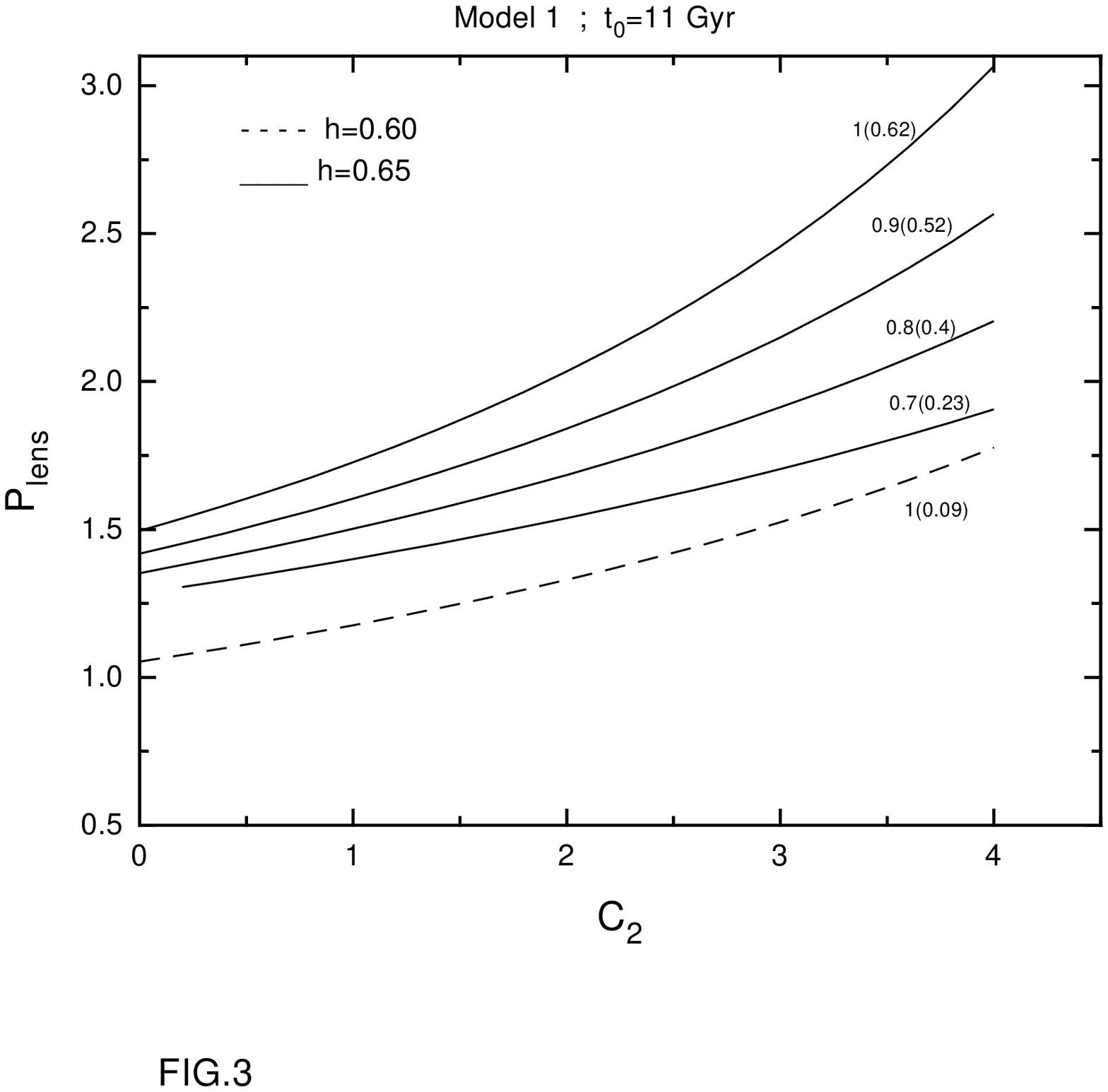}
\caption{Plot of $P_{lens}$ versus $C_2$ in Model 1 for an 
age of $t_0=11Gyr$. The first and second numbers on the curves correspond 
to $\Omega_M$ and $C_1$.}
\end{figure}
\begin{figure}
\epsscale{.8}
\plotone{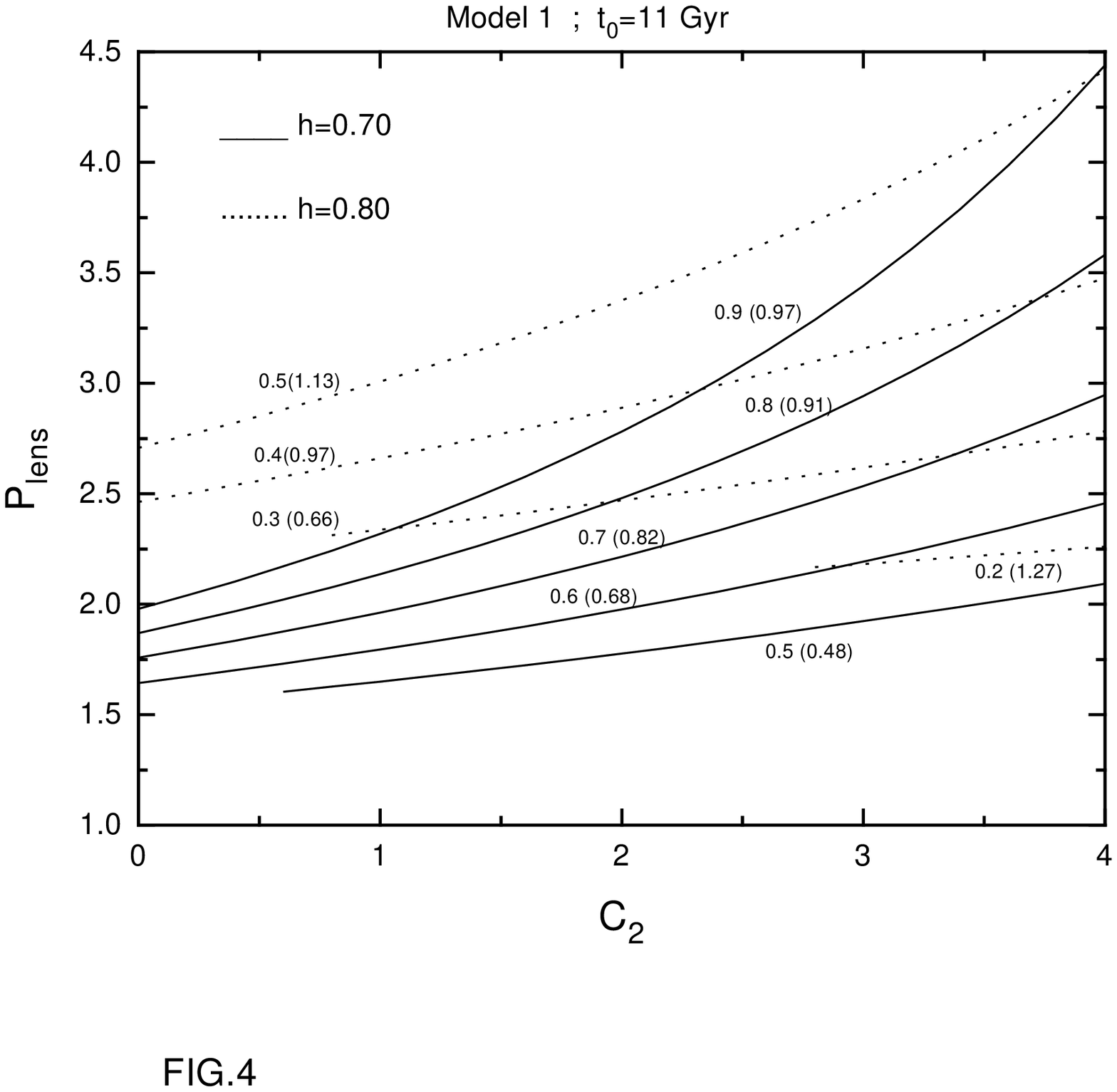}
\caption{Plot of $P_{lens}$ versus $C_2$ in Model 1 for an 
age of $t_0=11Gyr$. The first and second numbers on the curves correspond 
to $\Omega_M$ and $C_1$.}
\end{figure}
\begin{figure}
\epsscale{.8}
\plotone{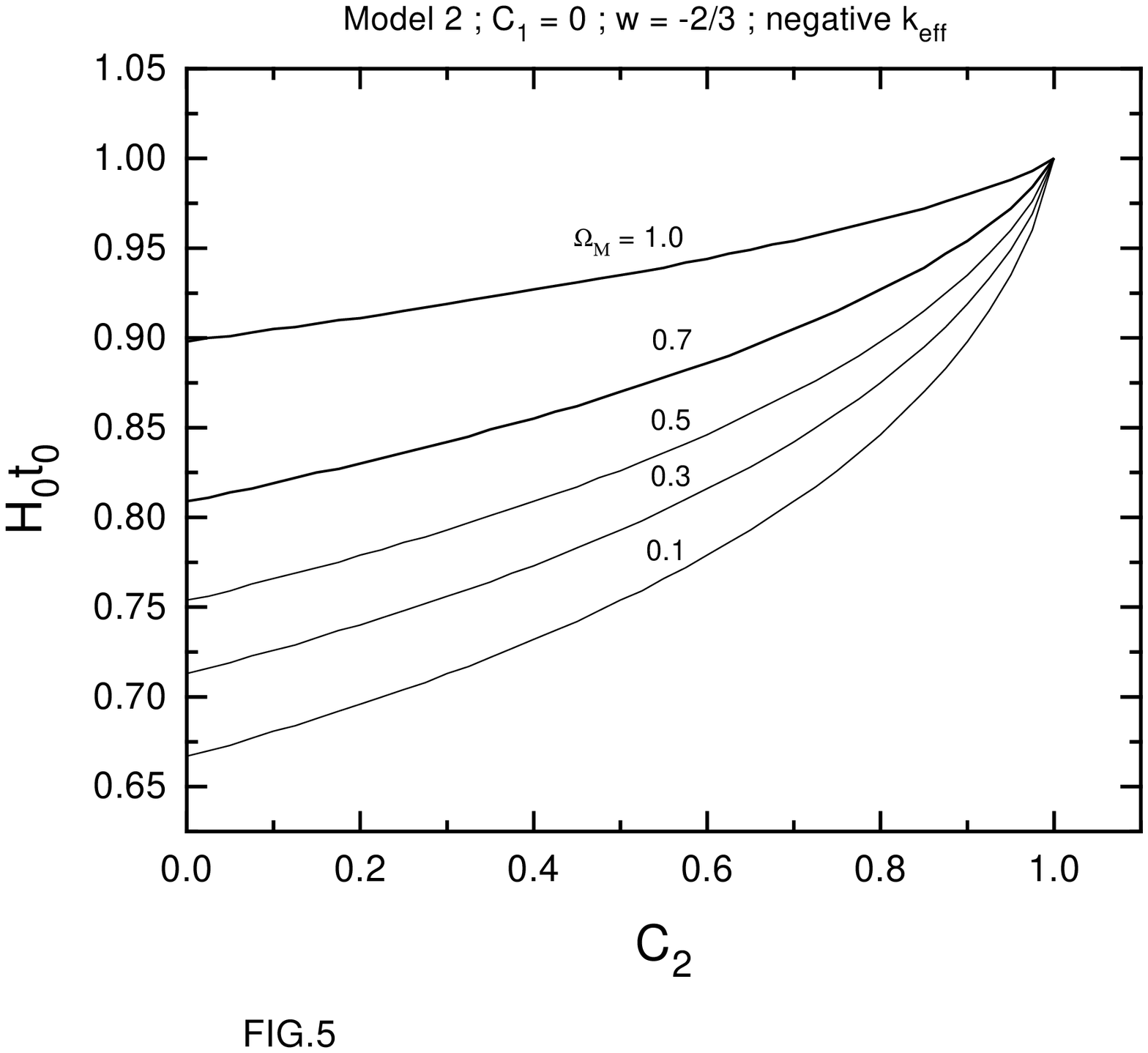}
\caption{Plot of $H_0t_0$ versus $C_2$ in Model 2 for 
various values of $\Omega_M$.}
\end{figure}
\end{document}